\documentclass{PoS}

\let\ifpdf\relax

\usepackage{amssymb}
\usepackage{graphicx}
\usepackage{amsmath}
\usepackage{amsfonts}
\usepackage{subfigure}
\usepackage{wrapfig}
\usepackage{epsfig}
\usepackage{afterpage,float}
\usepackage{cite}
\usepackage{ifpdf}

\title{Neutron-antineutron oscillations on the lattice}

\ShortTitle{Neutron-antineutron oscillations on the lattice}

\author{\speaker{Michael I. Buchoff}%
        \thanks{This work performed under the auspices of the U.S. Department of Energy by Lawrence Livermore National Laboratory under Contract DE-AC52-07NA27344. This work was partially supported by LDRD 10-ERD-033 and computing provided by LLNL Institutional Computing program.}
         \thanks{Preprint numbers:  LLNL-PROC-563736}, Chris Schroeder, Joseph Wasem\\
       Physical Sciences Directorate, Lawrence Livermore National Laboratory \\Livermore, California 94550, USA\\
       E-mail: \email{buchoff1@llnl.gov}}

\abstract{One possible low energy process due to beyond the Standard Model (BSM) physics is the neutron-antineutron transition, where baryon number changes by two units. In addition to providing a source of baryon number violation in the early universe, interactions of this kind are natural in grand unified theories (GUTs) with Majorana neutrinos that violate lepton number. Bounds on these oscillations can greatly restrict a variety of GUTs, while a non-zero signal would be a ``smoking gun" for new physics; however, to make a reliable prediction, the six-quark nucleon-antinucleon matrix elements must first be calculated non-perturbatively via lattice QCD.  We review the current understanding of this quantity, describe the lattice formalism, and present preliminary results from $32^3\times256$  clover-Wilson lattices with a pion mass of 390 MeV.
}

\FullConference{The XXX International Symposium on Lattice Field Theory\\
                June 24-29, 2012\\
                Cairns,  Australia}

\begin{document}
\def\a{{\alpha}}
\def\b{{\beta}}
\def\d{{\delta}}
\def\D{{\Delta}}
\def\X{{\Xi}}
\def\e{{\varepsilon}}
\def\g{{\gamma}}
\def\G{{\Gamma}}
\def\k{{\kappa}}
\def\l{{\lambda}}
\def\L{{\Lambda}}
\def\m{{\mu}}
\def\n{{\nu}}
\def\o{{\omega}}
\def\O{{\Omega}}
\def\S{{\Sigma}}
\def\s{{\sigma}}
\def\th{{\theta}}

\def\ol#1{{\overline{#1}}}

\def\Aslash{A\hskip-0.45em /}
\def\Dslash{D\hskip-0.65em /}
\def\Dtslash{\tilde{D} \hskip-0.65em /}

\def\CPT{{$\chi$PT}}
\def\QCPT{{Q$\chi$PT}}
\def\PQCPT{{PQ$\chi$PT}}
\def\tr{\text{tr}}
\def\str{\text{str}}
\def\diag{\text{diag}}
\def\order{{\mathcal O}}

\def\cF{{\mathcal F}}
\def\cS{{\mathcal S}}
\def\cC{{\mathcal C}}
\def\cB{{\mathcal B}}
\def\cT{{\mathcal T}}
\def\cQ{{\mathcal Q}}
\def\cL{{\mathcal L}}
\def\cO{{\mathcal O}}
\def\cA{{\mathcal A}}
\def\cQ{{\mathcal Q}}
\def\cR{{\mathcal R}}
\def\cH{{\mathcal H}}
\def\cW{{\mathcal W}}
\def\cM{{\mathcal M}}
\def\cD{{\mathcal D}}
\def\cN{{\mathcal N}}
\def\cP{{\mathcal P}}
\def\cK{{\mathcal K}}
\def\Qt{{\tilde{Q}}}
\def\Dt{{\tilde{D}}}
\def\St{{\tilde{\Sigma}}}
\def\cBt{{\tilde{\mathcal{B}}}}
\def\cDt{{\tilde{\mathcal{D}}}}
\def\cTt{{\tilde{\mathcal{T}}}}
\def\cMt{{\tilde{\mathcal{M}}}}
\def\At{{\tilde{A}}}
\def\cNt{{\tilde{\mathcal{N}}}}
\def\cOt{{\tilde{\mathcal{O}}}}
\def\cPt{{\tilde{\mathcal{P}}}}
\def\cI{{\mathcal{I}}}
\def\cJ{{\mathcal{J}}}

\def\eqref#1{{(\ref{#1})}}

\section{Introduction}

One unanswered mystery of the universe is the process that led to the abundance of observed baryons as compared to their antibaryon counterparts.  The source of this baryon number violation, which is expected to come from beyond the Standard Model (BSM) physics, can be realized in low-energy processes such as proton decay (if baryon number is violated by 1 unit) or transitions between neutrons and antineutrons (if baryon number is violated by 2 units).    The latter case, often referred to as neutron-antineutron oscillation (akin to neutral meson mixing), proves to be an intriguing scenario when considering the usual sphaleron picture of baryogenisis (which violates baryon number, $B$, and lepton number, $L$, but conserves $B-L$) coupled with Majorana neutrinos \cite{Mohapatra:1980qe} (whose transition between neutrinos and antineutrinos leads to $\Delta L = 2$).  Additionally, neutron-antineutron oscillations do not suffer from kinematic suppressions that can restrict proton decay if there is little overlap with the initial state proton and final state electron or muon \cite{Nussinov:2001rb}.  To that end, neutron-antineutron oscillations have been explored experimentally with intriguing prospects for $\mathcal{O}(1000)$ improvements in upcoming experimental efforts \cite{Proj_X}.

Any discussion of neutron-antineutron oscillations starts with assuming the existence of some BSM process that leads to a $\Delta B = 2$ operator in the low-energy effective field theory.  This operator will lead to off-diagonal elements of the Hamiltonian of the neutron-antineutron system
\begin{eqnarray}
H = \begin{pmatrix} E_n & \delta m \\ \delta m & E_{\bar{n}} \end{pmatrix}=\begin{pmatrix} E+V & \delta m \\ \delta m & E-V \end{pmatrix},
\end{eqnarray}
where $V$ is the potential difference between the neutron and the antineutron (a magnetic field can lead to a non-zero $V$ since the magnetic moments have opposite signs) and $V=0$ in a free system.  Upon solving the Schr\"odinger equation for the system, one finds the transition probability between neutrons and antineutrons given by
\begin{equation}
P_{n\rightarrow \bar{n}}(t) = \frac{\delta m^2}{\delta m^2 + V^2}\sin^2\Big[\sqrt{\delta m^2 + V^2} \ t \Big].
\end{equation}
While this equation is true for a given $V$, it is standard to define the period of free neutron oscillations due to the BSM physics as
\begin{equation}
\tau_{n\overline{n}} = \frac{1}{\delta m}.
\end{equation}
The value for $\tau_{n\overline{n}}$ greatly depends on which BSM scenario is being explored.   It has been estimated that a bound of $\tau_{n\overline{n}} \gtrsim 10^{10}-10^{11}\  \text{seconds}$ is sufficient to rule out many of the current models\cite{Mohapatra:2009wp}.   For example, TeV-scale seesaw mechanisms for neutrino masses in $SU(2)_L \times SU(2)_R \times SU(4)_c$ are expected to be ruled out at $\tau_{n\overline{n}} \gtrsim 10^{10}-10^{11}\  \text{seconds}$ \cite{Babu:2008rq}; and $SO(10)$ seesaw mechanisms with adequate baryogenisis, at $\tau_{n\overline{n}} \gtrsim 10^{9}-10^{12}\  \text{seconds}$ \cite{Babu:2012pp}.  Current experimental limits can even restrict extra-dimensional models with new particles with masses below a TeV \cite{Nussinov:2001rb,Winslow:2010wf}.  It should be emphasized that these are order of magnitude estimates with the QCD input coming from na\"ive dimensional analysis.  Future estimates will require rigorous and precise lattice calculations to keep pace with experimental precision. 

The  detection mechanism for these transitions is the cold annihilation of a newly formed antineutron with a nearby neutron (for more details, see W. M. Snow's plenary at PXPS 2012 \cite{Proj_X}).  The primary channel for this cold annihilation is $n\overline{n} \rightarrow 5\pi$, and this unique signature allows for experimental signals with little or no background.  Generally, there are two sets of experimental searches.  The first, which comes for free with large proton decay detectors such as Super-K, is based on neutron-antineutron annihilation within nuclei.  Na\"ively, one might expect this to occur quite frequently, as the number of nuclei far exceeds the expected bound $\tau_{n\overline{n}} \gtrsim 10^{11}$; however, the oscillation period within nuclei is highly suppressed, with a magnitude of roughly \cite{Friedman:2008es}
\begin{equation}
\tau_{Nucl}=(3 \times 10^{22})\frac{\tau_{n\overline{n}}^2}{\text{sec}}.
\end{equation}
As a result, the bound is suppressed compared to the free expectation, and one must rely on model estimations and extrapolations to extract it.  To date, the most stringent bound from experiments of this kind, $\tau_{n \overline{n}} > 3.5 \times 10^8 \ \text{seconds}$,  comes from Super-K (2011) \cite{Super-K}.

The second type of experiment explores the annihilation of free, cold neutrons with a target after a significant time of flight.  This type of experiment is free of the model-dependent estimations required for annihilations within nuclei and allows for greater control of systematics.  To date, the most stringent bound comes from the ILL experiment (1993): $\tau_{n \overline{n}} > 0.86 \times 10^8 \ \text{seconds}$ \cite{BaldoCeolin:1994jz}.  A factor of $\mathcal{O}(1000)$ increase is estimated for future experiments of this kind, but the bounds to rule out various BSM theories could be altered significantly depending on QCD enhancement or suppression of the neutron-antineutron matrix elements.

\section{Oscillations and matrix elements}
The observed value of the mixing arises from three inputs
\begin{equation}
\frac{1}{\tau_{n\overline{n}}}=\delta m = c_{BSM}(\mu_{BSM},\mu_W)c_{QCD}(\mu_W,\Lambda_{QCD}) \langle \overline{n}| \mathcal{O} | n \rangle,
\end{equation}
where $c_{BSM}$ is the running of the BSM theory to the weak interaction scale, $c_{QCD} $ is the QCD running from the weak to the nuclear scale, and  $\langle \overline{n}| \mathcal{O} | n \rangle$ is the non-perturbative matrix element mixing the neutron and antineutron states.  
The one-loop perturbative QCD running, $c_{QCD}$, is known \cite{Winslow:2010wf,Ozer:1982qh}, and $c_{BSM}$ has been calculated for multiple theories \cite{Nussinov:2001rb,Babu:2008rq,Babu:2012pp,Winslow:2010wf}.
The operator $\mathcal{O}$  contains two up quarks and four down quarks and is composed of three pairs of quarks from the possible forms
\begin{equation}
u^T C u \quad , \quad u^T C d \quad , \quad d^T C d \quad,
\end{equation}
where $C$ is the charge conjugation matrix.  Additionally, these terms always come in chiral pairs,
\begin{equation}
u^T_L C d_L \quad , \quad u^T_R C d_R \quad,
\end{equation}
since the mixed chirality terms are zero.  Lastly, these operators are invariant under color symmetry, $SU(3)_c$, which leads to two color tensors 
\begin{eqnarray}
\Gamma^{s}_{ijklmn}&=& \epsilon_{mik}\epsilon_{njl}+\epsilon_{nik}\epsilon_{mjl}+\epsilon_{mjk}\epsilon_{nil}+\epsilon_{njk}\epsilon_{mil} ,\quad\Gamma^{a}_{ijklmn}= \epsilon_{mij}\epsilon_{nkl}+\epsilon_{nij}\epsilon_{mkl},
\end{eqnarray}
where $i,j,k,l,m,n$ are color indices.  These three conditions lead to three types of operators \cite{Rao:1982gt}:
\begin{eqnarray}
\mathcal{O}^1_{\chi_1\chi_2\chi_3}&=&(u_{i\chi_1}^{T} C u_{j\chi_1})(d_{k\chi_2}^{T} C d_{l\chi_2})(d_{m\chi_3}^T C d_{n\chi_3})\Gamma^{s}_{ijklmn},\nonumber\\
\mathcal{O}^2_{\chi_1\chi_2\chi_3}&=&(u_{i\chi_1}^{T} C d_{j\chi_1})(u_{k\chi_2}^{T} C d_{l\chi_2})(d_{m\chi_3}^T C d_{n\chi_3})\Gamma^{s}_{ijklmn},\nonumber\\
\mathcal{O}^3_{\chi_1\chi_2\chi_3}&=&(u_{i\chi_1}^{T} C d_{j\chi_1})(u_{k\chi_2}^{T} C d_{l\chi_2})(d_{m\chi_3}^T C d_{n\chi_3})\Gamma^{a}_{ijklmn},
\end{eqnarray}
where $\chi_i=L,R$.  At first glance, there would appear to be 24 independent operators, but there are several additional symmetries.  The first set of symmetries due to the flavor structure is
\begin{eqnarray}\label{Sym_1}
\mathcal{O}^1_{\chi_1 L R}&=&\mathcal{O}^1_{\chi_1 R L},\quad \mathcal{O}^{2,3}_{L R \chi_3}=\mathcal{O}^{2,3}_{R L \chi_3},
\end{eqnarray}
which reduces the set to 18 independent operators.   An additional symmetry that emerges from antisymmetrizing pairs of epsilon tensors over four indices leads to (with $\sigma=L,R$ and $\rho=L,R$)\cite{Caswell:1982qs}
\begin{equation}\label{Sym_2}
\mathcal{O}^2_{\sigma \sigma \rho} - \mathcal{O}^1_{\sigma \sigma \rho} = 3\mathcal{O}^3_{\sigma \sigma \rho}, 
\end{equation}
which reduces the set to 14 operators.  In addition to enforcing $SU(3)_c$, it is also expected that the operators should be invariant under $SU(2)_L \otimes U(1)_Y$.  This gauge symmetry and the symmetries in Eq.~\eqref{Sym_1} leave only six operators:
\begin{eqnarray}\
\mathcal{P}_1&=&\mathcal{O}^1_{R R R},\quad\mathcal{P}_2=\mathcal{O}^2_{R R R},\quad\mathcal{P}_3=\mathcal{O}^3_{R R R},\nonumber\\
\quad\mathcal{P}_4&=&2\mathcal{O}^3_{L R R},\quad\mathcal{P}_5=4\mathcal{O}^3_{L L R},\quad\mathcal{P}_6=4(\mathcal{O}^1_{L L R}-\mathcal{O}^2_{L L R}).
\end{eqnarray}
We will present results for these operators; however, including the symmetry in Eq.~\eqref{Sym_2} leads to the conditions,
\begin{eqnarray}\label{check}
\mathcal{P}_6&=&-3\mathcal{P}_5,\quad\quad\mathcal{P}_2-\mathcal{P}_1=3\mathcal{P}_3,
\end{eqnarray}
which reduces the number of independent operators to four.  We will use these last two equations for a consistency check of our calculation.

\section{Lattice formalism and contraction details}
The mechanism to extract the neutron-antineutron matrix elements follows the common practice of taking ratios of three-point to two-point correlation functions.  In particular, the three correlation functions of interest and their large Euclidean time behavior are given by
\begin{eqnarray}
C_{NN}(t) = \langle N(t) \overline{N}(0) \rangle &\rightarrow& |\langle N|n\rangle |^2 e^{-m_n t}, \quad C_{\overline{N}\overline{N}}(t) = \langle \overline{N}(t) N(0) \rangle \rightarrow |\langle \overline{N}|\overline{n}\rangle |^2 e^{-m_n t}, \nonumber\\
C_{\overline{N}\mathcal{O}N}(t_1,t_2) &=& \langle N(t_2)\mathcal{O}(0) \overline{N}(-t_1) \rangle \rightarrow \langle \overline{n}|\overline{N}\rangle \langle N|n\rangle e^{-m_n (t_1+t_2)} \langle n |\mathcal{O}|\overline{n}\rangle.
\end{eqnarray}
The desired quantity of interest, $\langle n |\mathcal{O}|\overline{n}\rangle$, is the long time asymptote of a combination of these correlation functions,
\begin{equation}\label{ratio}
\mathcal{R}=\frac{C_{\overline{N}\mathcal{O}N}(t_1,t_2)}{C_{\overline{NN}}(t_1+t_2)}\Bigg[\frac{C_{NN}(t_1)C_{\overline{NN}}(t_2)C_{\overline{NN}}(t_1+t_2)}{C_{\overline{NN}}(t_1)C_{NN}(t_2)C_{NN}(t_1+t_2)}\Bigg]^\frac{1}{2}\rightarrow \langle \overline{n}| \mathcal{O} | n \rangle.
\end{equation}

\begin{figure}[t] 
   \centering
   \includegraphics[width=3in]{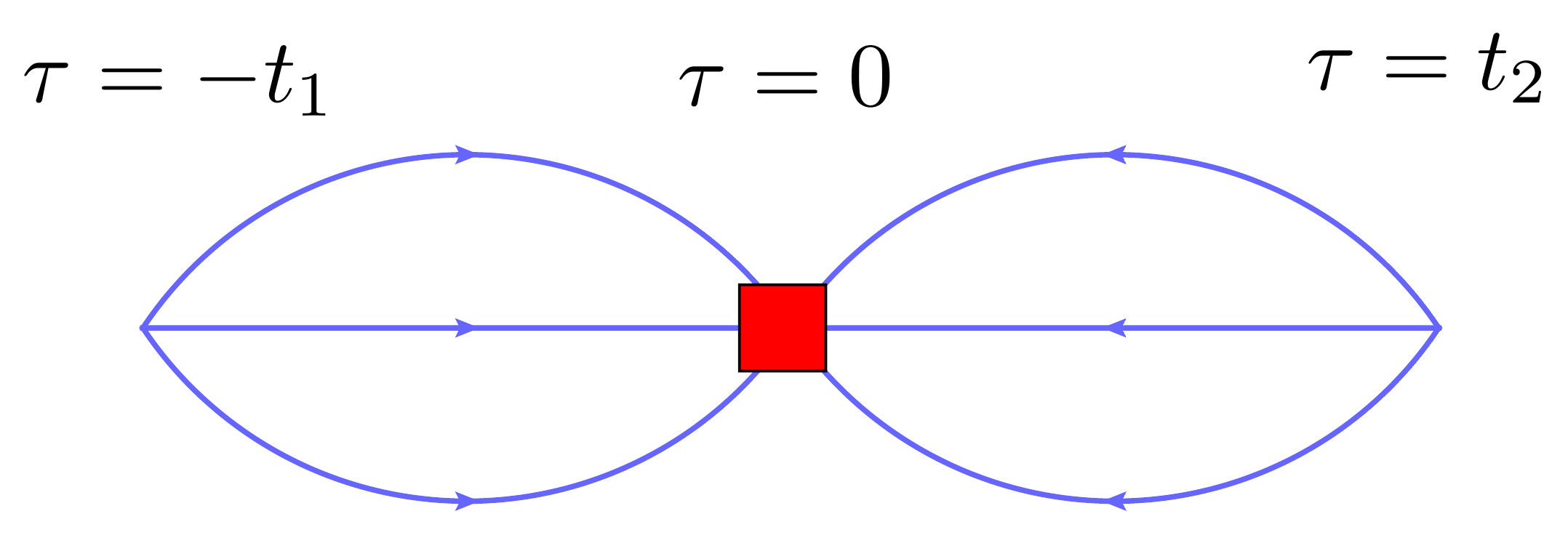} \quad \quad\quad\quad\includegraphics[width=2.1in]{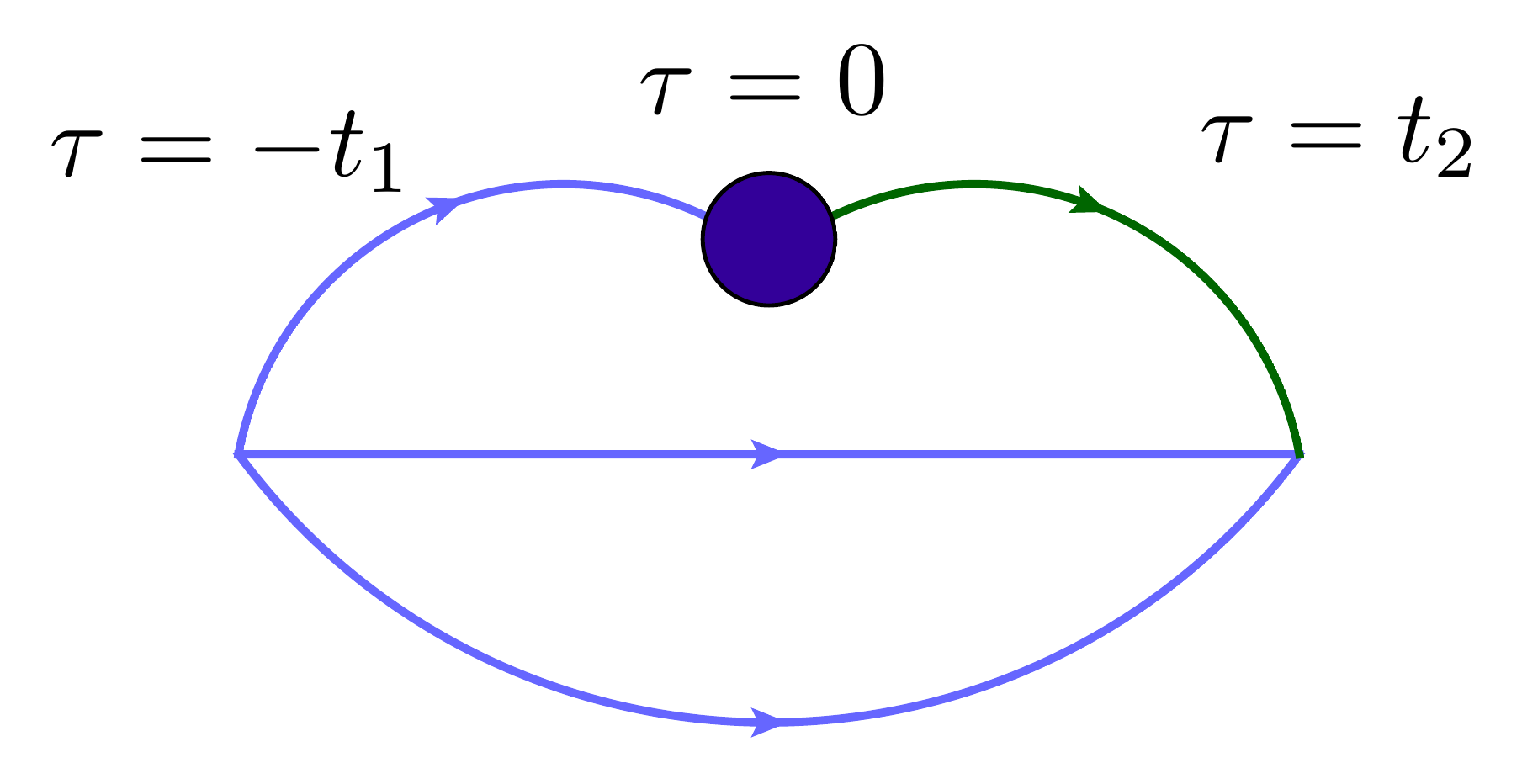}
   \caption{Comparison of neutron-antineutron three-point contractions (left) to typical bilinear three-point contractions (right).  One propagator is required for a measurements at all $(t_1,t_2)$ for the left diagram and two propagators are required for one measurement at a single $t_1$-value on the right diagram.}
   \label{fig:contract}
\end{figure}

The six-quark neutron-antineutron three-point correlation function has several key advantages over typical bi-linear or four-quark nucleon operators.  First, if the starting point for the propagator is at the operator insertion (as shown in Fig.~\ref{fig:contract}), only one propagator is needed per measurement,  whereas the typical nucleon three-point function requires two propagators, one that starts from the source and one that starts from the operator.  Second, because the propagator starts at the operator, one can acquire all the source-operator separations (given by $t_1$ in Fig.~\ref{fig:contract}) and operator-sink separations (given by $t_2$ in Fig.~\ref{fig:contract}), which allows for a two-dimensional analysis to quantify the excited state effect.  Alternatively, typical three point functions require far more computational resources to quantify excited state effects.  Lastly, the neutron-antineutron matrix element contains no disconnected or quark loop contractions, which removes the need for costly all-to-all propagators.  

One possible disadvantage  is that multiplying six propagators together (as done for the neutron-antineutron correlator) could increase the signal-to-noise degradation as compared to bilinear or four-quark matrix elements; however, we find a reasonably good signal-to-noise ratio, as shown in Fig.~\ref{fig:R_Plots}.

\section{Lattice Details}
The lattice calculations were performed with Chroma \cite{Edwards:2004sx} using the $32^3 \times 256$  anisotropic clover-Wilson lattices defined in Ref.~\cite{Lin:2008pr} with a pion mass of 390 MeV.  The temporal and spatial lattice spacings are roughly 0.035 and 0.123 fm, respectively, and the total spatial extent is roughly 4 fm ($m_\pi L \sim  7.8$).  For this preliminary calculation, we use a total of 159 configurations, each separated by 4 trajectories, to calculate 7268 propagators with Gaussian-smeared sources.   Contractions of these propagators lead to the same number of measurements at all source-operator and operator-sink separations.

\section{Preliminary results}
\begin{figure}[t] 
   \centering
   \includegraphics[width=5in]{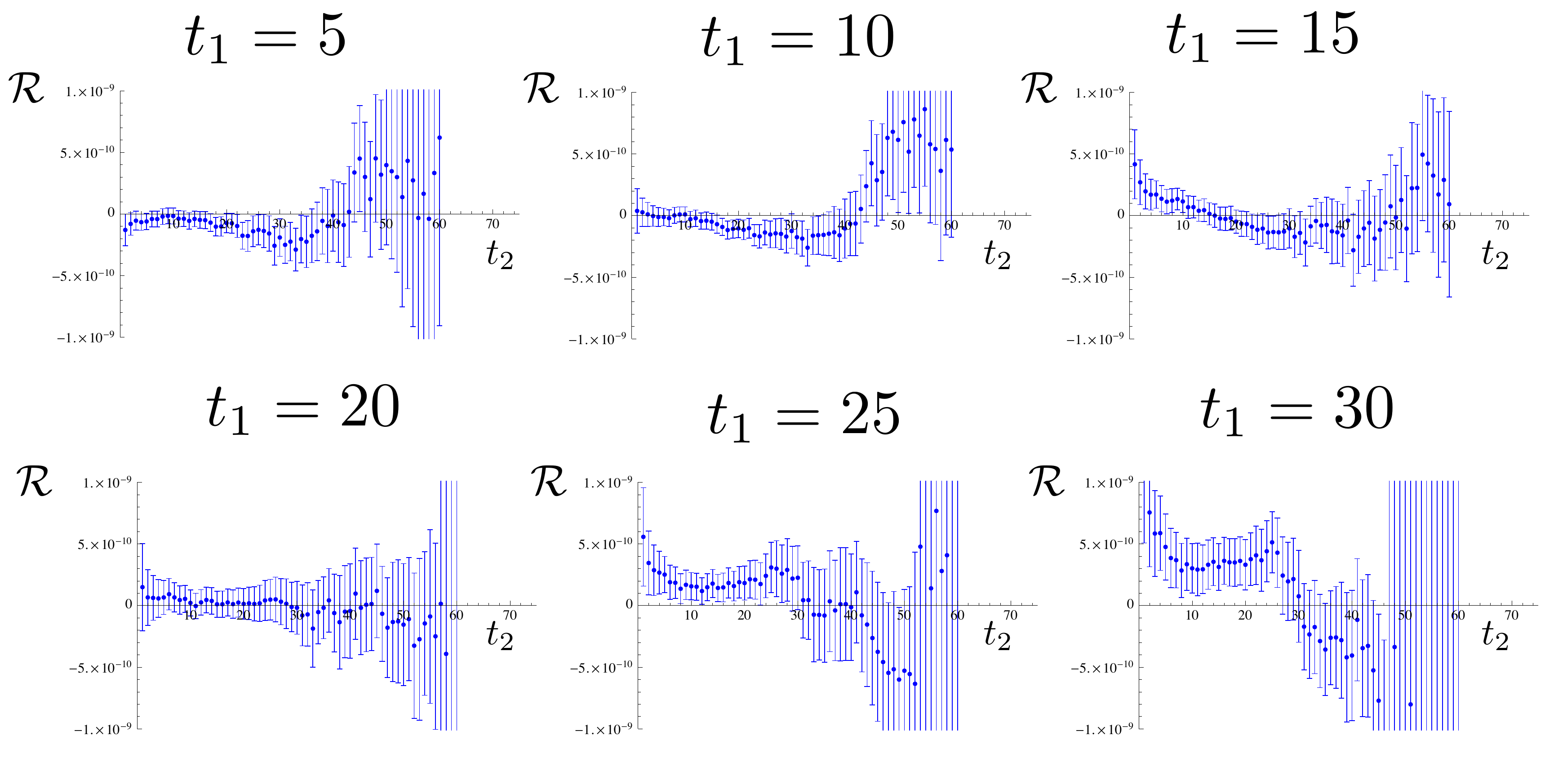}
   \caption{Plots of $\mathcal{R}$ vs. $t_2$ for six values of $t_1=5,10,15,20,25,30$.  The large $t_1$ and $t_2$ behavior of $\mathcal{R}$ should approach the neutron-antineutron matrix elements of interest.  For $t_1>30$ ($t_1\gtrsim1\ \text{fm}$), the plateau does not change appreciably, but signal-to-noise decreases.}
   \label{fig:R_Plots}
\end{figure}

\begin{figure}[b] 
   \centering
   \includegraphics[width=5.1in]{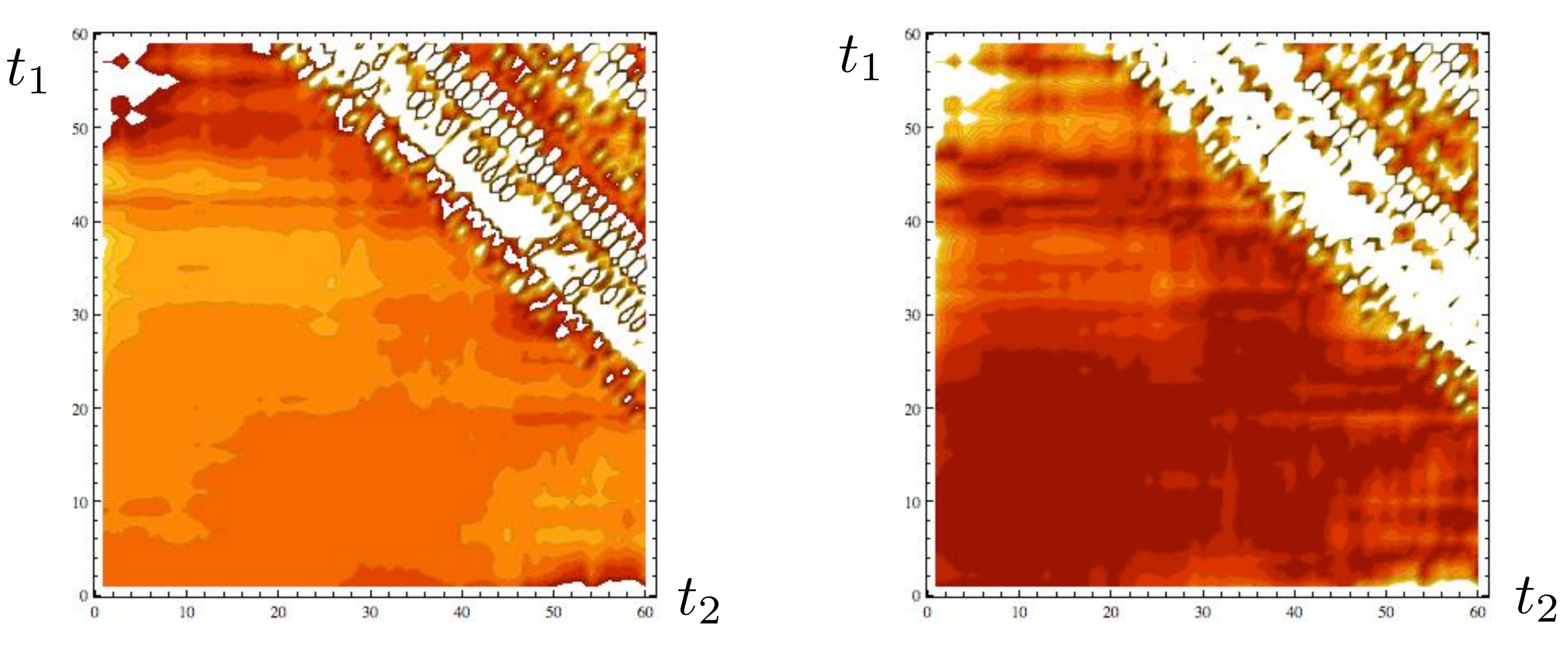} 
   \caption{Two-dimensional plot of $\mathcal{R}$ (left) and $|\mathcal{R}|$ (right) as a function of $t_1$ and $t_2$ ranging from time slices 0 to 60.  The lighter colors represent larger values of $\mathcal{R}$.  The two-dimensional plateau is achieved for $10<t_2<25$ and $30<t_1<40$.}
   \label{fig:2D_Plots}
\end{figure}

The desired matrix elements, $\langle \overline{n} | \mathcal{P}_i | n \rangle$, can be extracted from the long Euclidean time behavior of Eq.~\eqref{ratio}.  For each ratio $\mathcal{R}$, there are two time inputs, the source-operator separation ($t_1$) and the operator-sink separation ($t_2$).  In Fig.~\ref{fig:R_Plots}, $\mathcal{R}$ for the $\mathcal{P}_1$ operator  is plotted against $t_2$ for six different values of $t_1$.  Two features stand out from these plots.  First, there is a significant range of time slices where a signal can be extracted and the signal-to-noise degradation is not overly restrictive.  Second, it is evident that there is significant excited state dependence as $t_1$ is varied (for example, the plateaux extracted for $t_1 = 10$  and $t_1=30$ are significantly different).  For this reason, it is very important to use all information available to explore the full behavior of $\mathcal{R}$ as a function of both $t_1$ and $t_2$.

In Fig.~\ref{fig:2D_Plots}, the 2D plot of $\mathcal{R}$ and $|\mathcal{R}|$ are plotted against $t_1$ and $t_2$.  Again, it is clear that there is a significant amount of  non-trivial behavior due to excited states.  To that end, a 2D correlated fit has been performed over the time slices  $10<t_2<25$ and $30<t_1<40$.  For this preliminary calculation, systematic errors are estimated by adjusting 2D fit window  $\pm 1$ on all sides. 
\begin{table}[htbp]
   \centering
   \begin{tabular}{|c|c|c|} 
    \hline
      Operator    & Lattice Calculation ($10^{-5}\ \text{GeV}$) & MIT Bag Model Calculation ($10^{-5}\ \text{GeV}$)\\
      \hline
      $\langle \overline{n} | \mathcal{P}_1 | n \rangle$      & $1.57\pm 0.85^{+0.25}_{-0.30}$ & -6.56 \\
       \hline        
      $\langle \overline{n} | \mathcal{P}_2 | n \rangle$       & $-0.20\pm 0.14^{+0.14}_{-0.12}$  & 1.64 \\
       \hline
      $\langle \overline{n} | \mathcal{P}_3 | n \rangle$       & $-0.24\pm 0.26^{+0.10}_{-0.07}$  & 2.73 \\
       \hline
      $\langle \overline{n} | \mathcal{P}_4 | n \rangle$ &  $-0.02\pm 0.39^{+0.07}_{-0.18}$    &  -6.36 \\
       \hline
       $\langle \overline{n} | \mathcal{P}_5 | n \rangle$ & $0.34\pm 0.82^{+0.27}_{-0.57}$   &  9.64 \\
       \hline
       $\langle \overline{n} | \mathcal{P}_6 | n \rangle$ & $-2.07\pm 1.10^{+1.28}_{-0.77}$   &  -28.92 \\
       \hline
   \end{tabular}
   \caption{Table of results for neutron-antineutron matrix elements from the bare, non-renormalized lattice calculation and MIT bag model calculations \cite{Rao:1982gt}.} 
   \label{tab:results}
\end{table}

The bare (unrenormalized) results for $\langle \overline{n} | \mathcal{P}_i | n \rangle$ are shown in Table~\ref{tab:results}.  Eq.~\eqref{check} is satisfied exactly, configuration by configuration, but only stochastically in Table~\ref{tab:results} due to the bootstrapping in the analysis.
The corresponding values calculated from the MIT bag model are also displayed for comparison.  The magnitude of each operator as computed on the lattice is below that derived using the MIT bag model; however, it should be emphasized that the lattice results are very preliminary and still require renormalization factors.

\section{Systematic effects}
The primary systematic uncertainty in comparing the results of Table~\ref{tab:results} to experiment is the unphysically large pion mass used.
While it is not clear that an IR quantity such as the pion mass should dramatically effect the short distance six-quark vertex, it is a distinct possibility given that contractions of this system are similar to those for low-energy $NN$ scattering, where physical quark masses are expected to lead to a dramatic increase in the scattering length \cite{Beane:2006mx}.  

The second source of systematic uncertainty is the lattice cutoff (i.e., discretization) and matching the lattice regularization to the usual $\overline{MS}$ scheme used in the perturbative running \cite{Ozer:1982qh, Winslow:2010wf}.  Generically, operators of interest might mix with lower dimensional operators with the same symmetries, leading to diverging $1/a$ corrections (where $a$ is the lattice spacing).  However, this is not an issue for these operators since the lowest dimension operator that can lead to a $\Delta B=2$ interaction requires six quarks.  Regardless, there are expected to be $\mathcal{O}(a)$ corrections and renormalization coefficients that should be quantified.

The third systematic which was clear from Fig.~\ref{fig:R_Plots} and Fig.~\ref{fig:2D_Plots} is excited state contamination.  The calculation of the six-quark neutron-antineutron correlator gives us a unique view of these contaminations as a 2D function in $t_1$ and $t_2$, which is  difficult to come by for any other nucleon three-point function.  For this reason, we should be able to accurately quantify these contaminations.   Finally, finite volume effects should be quantified as well, though their impact is expected to be insignificant given the $m_\pi L \sim 7.8$ lattice used.

\section{Future prospects}
We are in the process of taking several steps to improve upon this very preliminary work: we are extending the calculation presented here as well as repeating it for a lighter, $240 \ \text{MeV}$ pion mass at the same volume and for the current pion mass with a  smaller, $2.5 \ \text{fm}$ spatial extent with a larger ensemble. We are exploring perturbative and non-perturbative lattice renormalization to properly match onto the perturbative QCD running previously calculated. We are also refining our analysis procedures to better quantify excited state effects.


Within the next year or two, we hope to carry out this calculation both with physical pion masses and with a chiral fermion discretization (domain-wall fermions).  Both calculations are numerically expensive, but within reach of the LLNL 20 PetaFlops Sequoia BG/Q.

\end{document}